# Target Atmospheric $CO_2$: Where Should Humanity Aim?

James Hansen*,[1,2], Makiko Sato[1,2], Pushker Kharecha[1,2], David Beerling[3], Robert Berner[4], Valerie Masson-Delmotte[5], Mark Pagani[4], Maureen Raymo[6], Dana L. Royer[7] and James C. Zachos[8]

[1]*NASA/Goddard Institute for Space Studies, New York, NY 10025, USA*
[2]*Columbia University Earth Institute, New York, NY 10027, USA*
[3]*Dept. Animal and Plant Sciences, University of Sheffield, Sheffield S10 2TN, UK*
[4]*Dept. Geology and Geophysics, Yale University, New Haven, CT 06520-8109, USA*
[5]*Lab. Des Sciences du Climat et l'Environnement/Institut Pierre Simon Laplace, CEA-CNRS-Universite de Versailles Saint-Quentin en Yvelines, CE Saclay, 91191, Gif-sur-Yvette, France*
[6]*Dept. Earth Sciences, Boston University, Boston, MA 02215, USA*
[7]*Dept. Earth and Environmental Sciences, Wesleyan University, Middletown, CT 06459-0139, USA*
[8]*Earth & Planetary Sciences Dept., University of California, Santa Cruz, Santa Cruz, CA 95064, USA*

**Abstract:** Paleoclimate data show that climate sensitivity is ~3°C for doubled $CO_2$, including only fast feedback processes. Equilibrium sensitivity, including slower surface albedo feedbacks, is ~6°C for doubled $CO_2$ for the range of climate states between glacial conditions and ice-free Antarctica. Decreasing $CO_2$ was the main cause of a cooling trend that began 50 million years ago, the planet being nearly ice-free until $CO_2$ fell to 450 ± 100 ppm; barring prompt policy changes, that critical level will be passed, in the opposite direction, within decades. If humanity wishes to preserve a planet similar to that on which civilization developed and to which life on Earth is adapted, paleoclimate evidence and ongoing climate change suggest that $CO_2$ will need to be reduced from its current 385 ppm to at most 350 ppm, but likely less than that. The largest uncertainty in the target arises from possible changes of non-$CO_2$ forcings. An initial 350 ppm $CO_2$ target may be achievable by phasing out coal use except where $CO_2$ is captured and adopting agricultural and forestry practices that sequester carbon. If the present overshoot of this target $CO_2$ is not brief, there is a possibility of seeding irreversible catastrophic effects.

## 1. INTRODUCTION

Human activities are altering Earth's atmospheric composition. Concern about global warming due to long-lived human-made greenhouse gases (GHGs) led to the United Nations Framework Convention on Climate Change [1] with the objective of stabilizing GHGs in the atmosphere at a level preventing "dangerous anthropogenic interference with the climate system."

The Intergovernmental Panel on Climate Change [IPCC, 2] and others [3] used several "reasons for concern" to estimate that global warming of more than 2-3°C may be dangerous. The European Union adopted 2°C above pre-industrial global temperature as a goal to limit human-made warming [4]. Hansen et al. [5] argued for a limit of 1°C global warming (relative to 2000, 1.7°C relative to pre-industrial time), aiming to avoid practically irreversible ice sheet and species loss. This 1°C limit, with nominal climate sensitivity of ¾°C per W/m$^2$ and plausible control of other GHGs [6], implies maximum $CO_2$ ~ 450 ppm [5].

Our current analysis suggests that humanity must aim for an even lower level of GHGs. Paleoclimate data and ongoing global changes indicate that 'slow' climate feedback processes not included in most climate models, such as ice sheet disintegration, vegetation migration, and GHG release from soils, tundra or ocean sediments, may begin to come into play on time scales as short as centuries or less [7]. Rapid on-going climate changes and realization that Earth is out of energy balance, implying that more warming is 'in the pipeline' [8], add urgency to investigation of the dangerous level of GHGs.

---

*Address correspondence to this author at NASA/Goddard Institute for Space Studies, New York, NY 10025, USA; E-mail: jhansen@giss.nasa.gov

A probabilistic analysis [9] concluded that the long-term $CO_2$ limit is in the range 300-500 ppm for 25 percent risk tolerance, depending on climate sensitivity and non-$CO_2$ forcings. Stabilizing atmospheric $CO_2$ and climate requires that net $CO_2$ emissions approach zero, because of the long lifetime of $CO_2$ [10, 11].

We use paleoclimate data to show that long-term climate has high sensitivity to climate forcings and that the present global mean $CO_2$, 385 ppm, is already in the dangerous zone. Despite rapid current $CO_2$ growth, ~2 ppm/year, we show that it is conceivable to reduce $CO_2$ this century to less than the current amount, but only via prompt policy changes.

## 1.1. Climate sensitivity

A global climate forcing, measured in W/m$^2$ averaged over the planet, is an imposed perturbation of the planet's energy balance. Increase of solar irradiance (So) by 2% and doubling of atmospheric $CO_2$ are each forcings of about 4 W/m$^2$ [12].

Charney [13] defined an idealized climate sensitivity problem, asking how much global surface temperature would increase if atmospheric $CO_2$ were instantly doubled, assuming that slowly-changing planetary surface conditions, such as ice sheets and forest cover, were fixed. Long-lived GHGs, except for the specified $CO_2$ change, were also fixed, not responding to climate change. The Charney problem thus provides a measure of climate sensitivity including only the effect of 'fast' feedback processes, such as changes of water vapor, clouds and sea ice.

Classification of climate change mechanisms into fast and slow feedbacks is useful, even though time scales of these changes may overlap. We include as fast feedbacks aerosol changes, e.g., of desert dust and marine dimethylsulfide, that occur in response to climate change [7].

Charney [13] used climate models to estimate fast-feedback doubled $CO_2$ sensitivity of 3 ± 1.5°C. Water vapor increase and sea ice decrease in response to global warming were both found to be strong positive feedbacks, amplifying the surface temperature response. Climate models in the current IPCC [2] assessment still agree with Charney's estimate.

Climate models alone are unable to define climate sensitivity more precisely, because it is difficult to prove that models realistically incorporate all feedback processes. The Earth's history, however, allows empirical inference of both fast feedback climate sensitivity and long-term sensitivity to specified GHG change including the slow ice sheet feedback.

## 2. PLEISTOCENE EPOCH

Atmospheric composition and surface properties in the late Pleistocene are known well enough for accurate assessment of the fast-feedback (Charney) climate sensitivity. We first compare the pre-industrial Holocene with the last glacial maximum [LGM, 20 ky BP (before present)]. The planet was in energy balance in both periods within a small fraction of 1 W/m$^2$, as shown by considering the contrary: an imbalance of 1 W/m$^2$ maintained a few millennia would melt all ice on the planet or change ocean temperature an amount far outside measured variations [Table **S1** of 8]. The approximate equilibrium characterizing most of Earth's history is unlike the current situation, in which GHGs are rising at a rate much faster than the coupled climate system can respond.

Climate forcing in the LGM equilibrium state due to the ice age surface properties, i.e., increased ice area, different vegetation distribution, and continental shelf exposure, was -3.5 ± 1 W/m$^2$ [14] relative to the Holocene. Additional forcing due to reduced amounts of long-lived

GHGs ($CO_2$, $CH_4$, $N_2O$), including the indirect effects of $CH_4$ on tropospheric ozone and stratospheric water vapor (Fig. **S1**) was -3 ± 0.5 W/m². Global forcing due to slight changes in the Earth's orbit is a negligible fraction of 1 W/m² (Fig. **S3**). The total 6.5 W/m² forcing and global surface temperature change of 5 ± 1°C relative to the Holocene [15, 16] yield an empirical sensitivity ~¾ ± ¼ °C per W/m² forcing, i.e., a Charney sensitivity of 3 ± 1 °C for the 4 W/m² forcing of doubled $CO_2$. This empirical fast-feedback climate sensitivity allows water vapor, clouds, aerosols, sea ice, and all other fast feedbacks that exist in the real world to respond naturally to global climate change.

Climate sensitivity varies as Earth becomes warmer or cooler. Toward colder extremes, as the area of sea ice grows, the planet approaches runaway snowball-Earth conditions, and at high temperatures it can approach a runaway greenhouse effect [12]. At its present temperature Earth is on a flat portion of its fast-feedback climate sensitivity curve (Fig. **S2**). Thus our empirical sensitivity, although strictly the mean fast-feedback sensitivity for climate states ranging from the ice age to the current interglacial period, is also today's fast-feedback climate sensitivity.

## 2.1. Verification

Our empirical fast-feedback climate sensitivity, derived by comparing conditions at two points in time, can be checked over the longer period of ice core data. Fig. **(1a)** shows $CO_2$ and $CH_4$ data from the Antarctic Vostok ice core [17, 18] and sea level based on Red Sea sediment cores [18]. Gases are from the same ice core and have a consistent time scale, but dating with respect to sea level may have errors up to several thousand years.

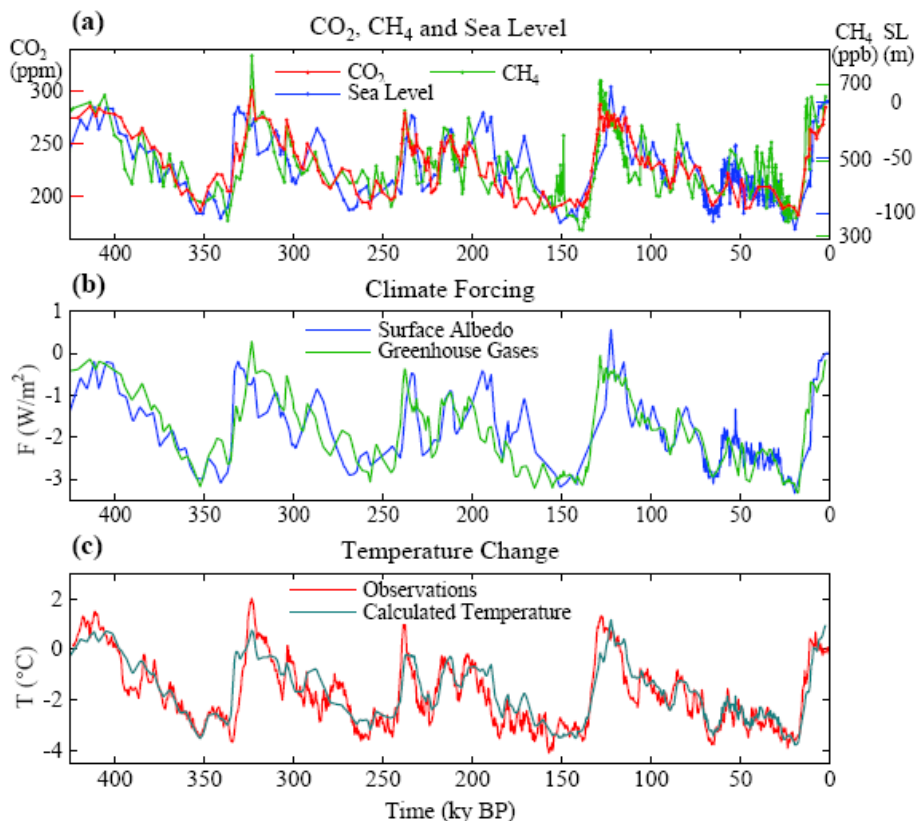

**Fig. (1).** (a) $CO_2$, $CH_4$ [17] and sea level [19] for past 425 ky.

(b) Climate forcings due to changes of GHGs and ice sheet area, the latter inferred from sea level change.

(c) Calculated global temperature change based on climate sensitivity of ¾°C per W/m². Observations are Antarctic temperature change [18] divided by two.

We use the GHG and sea level data to calculate climate forcing by GHGs and surface albedo change as in prior calculations [7], but with two refinements. First, we specify the $N_2O$ climate forcing as 12 percent of the sum of the $CO_2$ and $CH_4$ forcings, rather than the 15 percent estimated earlier [7] Because N$_2$O data are not available for the entire record, and its forcing is small and highly correlated with CO$_2$ and CH$_4$, we take the GHG effective forcing as

$$Fe (GHGs) = 1.12 [Fa(CO_2) + 1.4 Fa(CH_4)], \qquad (1)$$

using published formulae for Fa of each gas [20]. The factor 1.4 accounts for the higher efficacy of CH$_4$ relative to CO$_2$, which is due mainly to the indirect effect of CH$_4$ on tropospheric ozone and stratospheric water vapor [12]. The resulting GHG forcing between the LGM and late Holocene is 3 W/m$^2$, apportioned as 75% CO$_2$, 14% CH$_4$ and 11% N$_2$O.

The second refinement in our calculations is to surface albedo. Based on models of ice sheet shape, we take the horizontal area of the ice sheet as proportional to the 4/5 power of volume. Fig. **(S4)** compares our present albedo forcing with prior use [7] of exponent 2/3, showing that this choice and division of the ice into multiple ice sheets has only a minor effect.

Multiplying the sum of GHG and surface albedo forcings by climate sensitivity ¾°C per W/m$^2$ yields the blue curve in Fig. **(1c)**. Vostok temperature change [17] divided by two (red curve) is used to crudely estimate global temperature change, as typical glacial-interglacial global annual-mean temperature change is ~5°C and is associated with ~10°C change on Antarctica [21]. Fig. **(1c)** shows that fast-feedback climate sensitivity ¾°C per W/m$^2$ (3°C for doubled CO$_2$) is a good approximation for the entire period.

## 2.2. Slow feedbacks

Let us consider climate change averaged over a few thousand years – long enough to assure energy balance and minimize effects of ocean thermal response time and climate change leads/lags between hemispheres [22]. At such temporal resolution the temperature variations in Fig. **(1)** are global, with high latitude amplification, being present in polar ice cores and sea surface temperature derived from ocean sediment cores (Fig. **S5**).

GHG and surface albedo changes are mechanisms causing the large global climate changes in Fig. **(1)**, but they do not initiate these climate swings. Instead changes of GHGs and sea level (a measure of ice sheet size) lag temperature change by several hundred years [6, 7, 23, 24].

GHG and surface albedo changes are positive climate feedbacks. Major glacial-interglacial climate swings are instigated by slow changes of Earth's orbit, especially the tilt of Earth's spin-axis relative to the orbital plane and the precession of the equinoxes that influences the intensity of summer insolation [25, 26]. Global radiative forcing due to orbital changes is small, but ice sheet size is affected by changes of geographical and seasonal insolation (e.g., ice melts at both poles when the spin-axis tilt increases, and ice melts at one pole when perihelion, the closest approach to the sun, occurs in late spring [7]. Also a warming climate causes net release of GHGs. The most effective GHG feedback is release of CO$_2$ by the ocean, due partly to temperature dependence of CO$_2$ solubility but mostly to increased ocean mixing in a warmer climate, which acts to flush out deep ocean CO$_2$ and alters ocean biological productivity [27].

GHG and surface albedo feedbacks respond and contribute to temperature change caused by any climate forcing, natural or human-made, given sufficient time. The GHG feedback is nearly linear in global temperature during the late Pleistocene [Fig. **(7)** of 6, 28]. Surface albedo feedback increases as Earth becomes colder and the area of ice increases. Climate sensitivity on

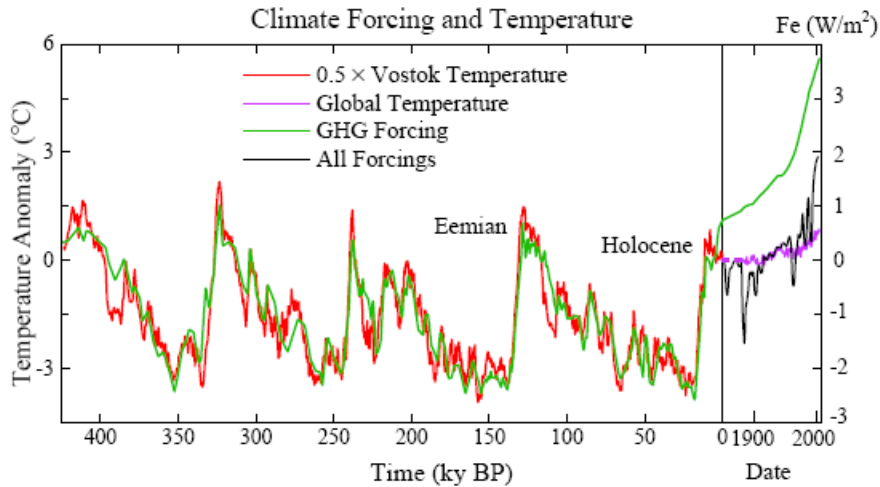

**Fig. (2).** Global temperature (left scale) and GHG forcing (right scale) due to $CO_2$, $CH_4$ and $N_2O$ from the Vostok ice core [17, 18]. Time scale is expanded for the industrial era. Ratio of temperature and forcing scales is 1.5°C per W/m$^2$, i.e., the temperature scale gives the expected equilibrium response to GHG change including (slow feedback) surface albedo change. Modern forcings include human-made aerosols, volcanic aerosols and solar irradiance [5]. GHG forcing zero point is the mean for 10-8 ky BP (Fig. **S6**). Zero point of modern temperature and net climate forcing was set at 1850 [5], but this is also the zero point for 10-8 ky BP, as shown by the absence of a trend in Fig. (**S6**) and by the discussion of that figure.

Pleistocene time scales includes slow feedbacks, and is larger than the Charney sensitivity, because the dominant slow feedbacks are positive. Other feedbacks, e.g., the negative feedback of increased weathering as $CO_2$ increases, become important on longer geologic time scales.

Paleoclimate data permit evaluation of long-term sensitivity to specified GHG change. We assume only that, to first order, the area of ice is a function of global temperature. Plotting GHG forcing [7] from ice core data [18] against temperature shows that global climate sensitivity including the slow surface albedo feedback is 1.5°C per W/m$^2$ or 6°C for doubled $CO_2$ (Fig. **2**), twice as large as the Charney fast-feedback sensitivity. Note that we assume the area of ice and snow on the planet to be predominately dependent on global temperature, but some changes of regional ice sheet properties occur as part of the Earth orbital climate forcing (see Appendix).

This equilibrium sensitivity of 6°C for doubled $CO_2$ is valid for specified GHG amount, as in studies that employ emission scenarios and coupled carbon cycle/climate models to determine GHG amount. If GHGs are included as a feedback (with say solar irradiance as forcing) sensitivity is still larger on Pleistocene time scales (see Appendix), but the sensitivity may be reduced by negative feedbacks on geologic time scales [29, 30]. The 6°C sensitivity reduces to 3°C when the planet has become warm enough to lose its ice sheets.

This long-term climate sensitivity is relevant to GHGs that remain airborne for centuries-to-millennia. The human-caused atmospheric GHG increase will decline slowly if anthropogenic emissions from fossil fuel burning decrease enough, as we illustrate below using a simplified carbon cycle model. On the other hand, if the globe warms much further, carbon cycle models [2] and empirical data [6, 28] reveal a positive GHG feedback on century-millennia time scales. This amplification of GHG amount is moderate if warming is kept within the range of recent interglacial periods [6], but larger warming would risk greater release of $CH_4$ and $CO_2$ from methane hydrates in tundra and ocean sediments [29]. On still longer, geological, time scales weathering of rocks causes a negative feedback on atmospheric $CO_2$ amount [30], as discussed in section 3, but this feedback is too slow to alleviate climate change of concern to humanity.

**2.3 Time scales**

How long does it take to reach equilibrium temperature with specified GHG change? Response is slowed by ocean thermal inertia and the time needed for ice sheets to disintegrate.

Ocean-caused delay is estimated in Fig. **(S7)** using a coupled atmosphere-ocean model. One-third of the response occurs in the first few years, in part because of rapid response over land, one-half in ~25 years, three-quarters in 250 years, and nearly full response in a millennium. The ocean-caused delay is a strong (quadratic) function of climate sensitivity and it depends on the rate of mixing of surface water and deep water [31], as discussed in the Appendix.

Ice sheet response time is often assumed to be several millennia, based on the broad sweep of paleo sea level change (Fig.**1a**) and primitive ice sheet models designed to capture that change. However, this long time scale may reflect the slowly changing orbital forcing, rather than inherent inertia, as there is no discernable lag between maximum ice sheet melt rate and local insolation that favors melt [7]. Paleo sea level data with high time resolution reveal frequent 'suborbital' sea level changes at rates of 1 m/century or more [32-34].

Present-day observations of Greenland and Antarctica show increasing surface melt [35], loss of buttressing ice shelves [36], accelerating ice streams [37], and increasing overall mass loss [38]. These rapid changes do not occur in existing ice sheet models, which are missing critical physics of ice sheet disintegration [39]. Sea level changes of several meters per century occur in the paleoclimate record [32, 33], in response to forcings slower and weaker than the present human-made forcing. It seems likely that large ice sheet response will occur within centuries, if human-made forcings continue to increase. Once ice sheet disintegration is underway, decadal changes of sea level may be substantial.

## 2.4. Warming "in the pipeline"

The expanded time scale for the industrial era (Fig. **2**) reveals a growing gap between actual global temperature (purple curve) and equilibrium (long-term) temperature response based on the net estimated climate forcing (black curve). Ocean and ice sheet response times together account for this gap, which is now 2.0°C.

The forcing in Fig. **(2)** (black curve, Fe scale), when used to drive a global climate model [5], yields global temperature change that agrees closely [Fig. **(3)** in 5] with observations (purple curve, Fig. **2**). That climate model, which includes only fast feedbacks, has additional warming of ~0.6°C in the pipeline today because of ocean thermal inertia [5, 8].

The remaining gap between equilibrium temperature for current atmospheric composition and actual global temperature is ~1.4°C. This further 1.4°C warming still to come is due to the slow surface albedo feedback, specifically ice sheet disintegration and vegetation change.

One may ask whether the climate system, as the Earth warms from its present 'interglacial' state, still has the capacity to supply slow feedbacks that double the fast-feedback sensitivity. This issue can be addressed by considering longer time scales including periods with no ice.

## 3. CENOZOIC ERA

Pleistocene atmospheric $CO_2$ variations occur as a climate feedback, as carbon is exchanged among surface reservoirs: the ocean, atmosphere, soils and biosphere. The most effective feedback is increase of atmospheric $CO_2$ as climate warms, the $CO_2$ transfer being mainly from ocean to atmosphere [27, 28]. On longer time scales the total amount of $CO_2$ in the surface reservoirs varies due to exchange of carbon with the solid earth. $CO_2$ thus becomes a primary agent of long-term climate change, leaving orbital effects as 'noise' on larger climate swings.

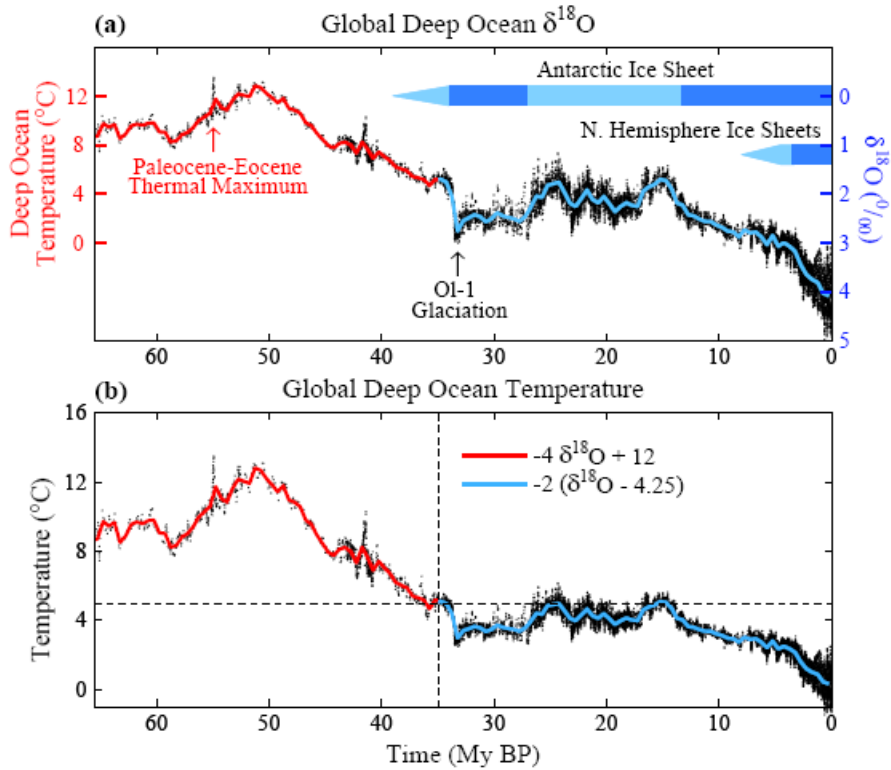

**Fig. (3).** Global deep ocean **(a)** $\delta^{18}O$ [26] and **(b)** temperature. Black curve is 5-point running mean of $\delta^{18}O$ original temporal resolution, while red and blue curves have 500 ky resolution.

The Cenozoic era, the past 65.5 My, provides a valuable complement to the Pleistocene for exploring climate sensitivity. Cenozoic data on climate and atmospheric composition are not as precise, but larger climate variations occur, including an ice-free planet, thus putting glacial-interglacial changes in a wider perspective.

Oxygen isotopic composition of benthic (deep ocean dwelling) foraminifera shells in a global compilation of ocean sediment cores [26] provides a starting point for analyzing Cenozoic climate change (Fig. **3a**). At times with negligible ice sheets, oxygen isotope change, $\delta^{18}O$, provides a direct measure of deep ocean temperature ($T_{do}$). Thus $T_{do}$ (°C) ~ $-4 \delta^{18}O + 12$ between 65.5 and 35 My BP.

Rapid increase of $\delta^{18}O$ at about 34 My is associated with glaciation of Antarctica [26, 40] and global cooling, as evidenced by data from North America [41] and Asia [42]. From then until the present, $^{18}O$ in deep ocean foraminifera is affected by both ice volume and $T_{do}$, lighter $^{16}O$ evaporating preferentially from the ocean and accumulating in ice sheets. Between 35 My and the last ice age (20 ky) the change of $\delta^{18}O$ was ~ 3‰, change of $T_{do}$ was ~ 6°C (from +5 to -1°C) and ice volume change ~ 180 msl (meters of sea level). Given that a 1.5‰ change of $\delta^{18}O$ is associated with a 6°C $T_{do}$ change, we assign the remaining $\delta^{18}O$ change to ice volume linearly at the rate 60 msl per mil $\delta^{18}O$ change (thus 180 msl for $\delta^{18}O$ between 1.75 and 4.75). Equal division of $\delta^{18}O$ between temperature and sea level yields sea level change in the late Pleistocene in reasonable accord with available sea level data (Fig. **S8**). Subtracting the ice volume portion of $\delta^{18}O$ yields deep ocean temperature $T_{do}$ (°C) = $-2 (\delta^{18}O -4.25‰)$ after 35 My, as in Fig. **(3b)**.

The large (~14°C) Cenozoic temperature change between 50 My and the ice age at 20 ky must have been forced by changes of atmospheric composition. Alternative drives could come from outside (solar irradiance) or the Earth's surface (continental locations). But solar brightness increased ~0.4% in the Cenozoic [43], a linear forcing change of only +1 W/m$^2$ and of the wrong sign to contribute to the cooling trend. Climate forcing due to continental locations was < 1 W/m$^2$, because continents 65 My ago were already close to present latitudes (Fig. **S9**). Opening or closing of oceanic gateways might affect the timing of glaciation, but it would not provide the climate forcing needed for global cooling.

$CO_2$ concentration, in contrast, varied from ~180 ppm in glacial times to 1500 ± 500 ppm in the early Cenozoic [44]. This change is a forcing of more than 10 W/m$^2$ (Table 1 in [16]), an order of magnitude larger than other known forcings. $CH_4$ and $N_2O$, positively correlated with $CO_2$ and global temperature in the period with accurate data (ice cores), likely increase the total GHG forcing, but their forcings are much smaller than that of $CO_2$ [45, 46].

**3.1. Cenozoic carbon cycle**

Solid Earth sources and sinks of $CO_2$ are not, in general, balanced at any given time [30, 47]. $CO_2$ is removed from surface reservoirs by: (1) chemical weathering of rocks with deposition of carbonates on the ocean floor, and (2) burial of organic matter; weathering is the dominant process [30]. $CO_2$ returns primarily via metamorphism and volcanic outgassing at locations where carbonate-rich oceanic crust is being subducted beneath moving continental plates.

Outgassing and burial of $CO_2$ are each typically $10^{12}$-$10^{13}$ mol C/year [30, 47-48]. At times of unusual plate tectonic activity, such as rapid subduction of carbon-rich ocean crust or strong orogeny, the imbalance between outgassing and burial can be a significant fraction of the one-way carbon flux. Although negative feedbacks in the geochemical carbon cycle reduce the rate of surface reservoir perturbation [49], a net imbalance ~$10^{12}$ mol C/year can be maintained over thousands of years. Such an imbalance, if confined to the atmosphere, would be ~0.005 ppm/year, but as $CO_2$ is distributed among surface reservoirs, this is only ~0.0001 ppm/year. This rate is negligible compared to the present human-made atmospheric $CO_2$ increase of ~2 ppm/year, yet over a million years such a crustal imbalance alters atmospheric $CO_2$ by 100 ppm.

Between 60 and 50 My ago India moved north rapidly, 18-20 cm/year [50], through a region that long had been a depocenter for carbonate and organic sediments. Subduction of carbon-rich crust was surely a large source of $CO_2$ outgassing and a prime cause of global warming, which peaked 50 My ago (Fig. **3b**) with the Indo-Asian collision. $CO_2$ must have then decreased due to a reduced subduction source and enhanced weathering with uplift of the Himalayas/Tibetan Plateau [51]. Since then, the Indian and Atlantic Oceans have been major depocenters for carbon, but subduction of carbon-rich crust has been limited mainly to small regions near Indonesia and Central America [47].

Thus atmospheric $CO_2$ declined following the Indo-Asian collision [44] and climate cooled (Fig. **3b**) leading to Antarctic glaciation by ~34 My. Antarctica has been more or less glaciated ever since. The rate of $CO_2$ drawdown declines as atmospheric $CO_2$ decreases due to negative feedbacks, including the effect of declining atmospheric temperature and plant growth rates on weathering [30]. These negative feedbacks tend to create a balance between crustal outgassing and drawdown of $CO_2$, which have been equal within 1-2 percent over the past 700 ky [52]. Large fluctuations in the size of the Antarctic ice sheet have occurred in the past 34 My, possibly related to temporal variations of plate tectonics [53] and outgassing rates. The relatively constant atmospheric $CO_2$ amount of the past 20 My (Fig. **S10**) implies a near balance of outgassing and weathering rates over that period.

Knowledge of Cenozoic $CO_2$ is limited to imprecise proxy measures except for recent ice core data. There are discrepancies among different proxy measures, and even between different investigators using the same proxy method, as discussed in conjunction with Fig. (**S10**). Nevertheless, the proxy data indicate that $CO_2$ was of the order of 1000 ppm in the early Cenozoic but <500 ppm in the last 20 My [2, 44].

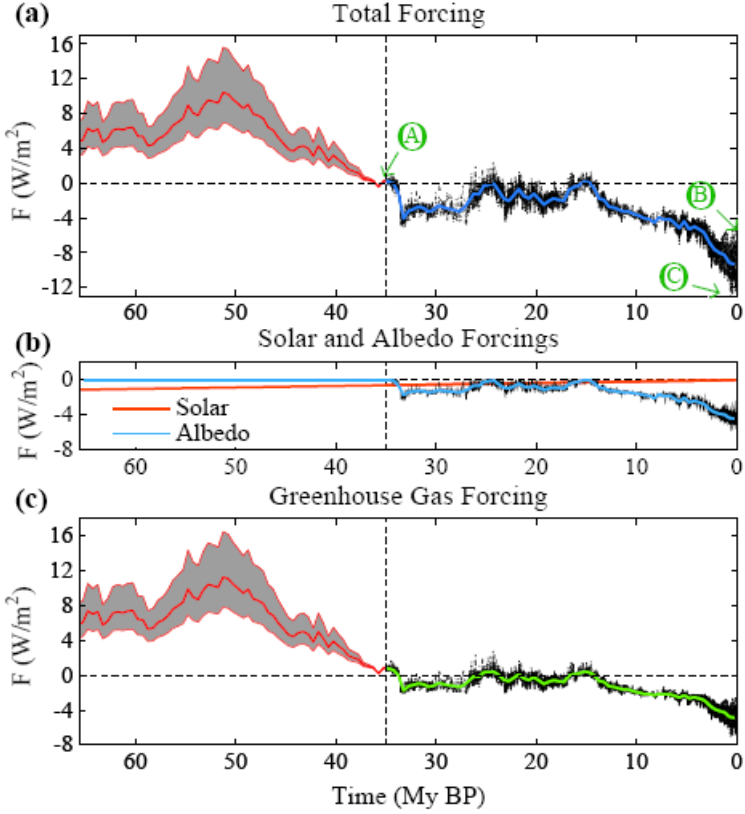

Fig. (4). (a) Total climate forcing, (b) solar and surface albedo forcings, and (c) GHG forcing in the Cenozoic, based on $T_{do}$ history of Fig. (3b) and assumed fast-feedback climate sensitivity ¾°C per W/m². Ratio of $T_s$ change and $T_{do}$ change is assumed to be near unity in the minimal ice world between 65 and 35 My, but the gray area allows for 50% uncertainty in the ratio. In the later era with large ice sheets we take $\Delta T_s/\Delta T_{do} = 1.5$, in accord with Pleistocene data.

### 3.2. Cenozoic forcing and CO₂

The entire Cenozoic climate forcing history (Fig. **4a**) is implied by the temperature reconstruction (Fig. **3b**), assuming a fast-feedback sensitivity of ¾°C per W/m². Subtracting the solar and surface albedo forcings (Fig. **4b**), the latter from Eq. S2 with ice sheet area vs. time from $\delta^{18}O$, we obtain the GHG forcing history (Fig. **4c**).

We hinge our calculations at 35 My for several reasons. Between 65 and 35 My ago there was little ice on the planet, so climate sensitivity is defined mainly by fast feedbacks. Second, we want to estimate the $CO_2$ amount that precipitated Antarctic glaciation. Finally, the relation between global surface air temperature change ($\Delta T_s$) and deep ocean temperature change ($\Delta T_{do}$) differs for ice-free and glaciated worlds.

Climate models show that global temperature change is tied closely to ocean temperature change [54]. Deep ocean temperature is a function of high latitude ocean surface temperature, which tends to be amplified relative to global mean ocean surface temperature. However, land temperature change exceeds that of the ocean, with an effect on global temperature that tends to offset the latitudinal variation of ocean temperature. Thus in the ice-free world (65-35 My) we take $\Delta T_s \sim \Delta T_{do}$ with generous (50%) uncertainty. In the glaciated world $\Delta T_{do}$ is limited by the freezing point in the deep ocean. $\Delta T_s$ between the last ice age (20 ky) and the present interglacial period (~5°C) was ~1.5 times larger than $\Delta T_{do}$. In Fig. (**S5**) we show that this relationship fits well throughout the period of ice core data.

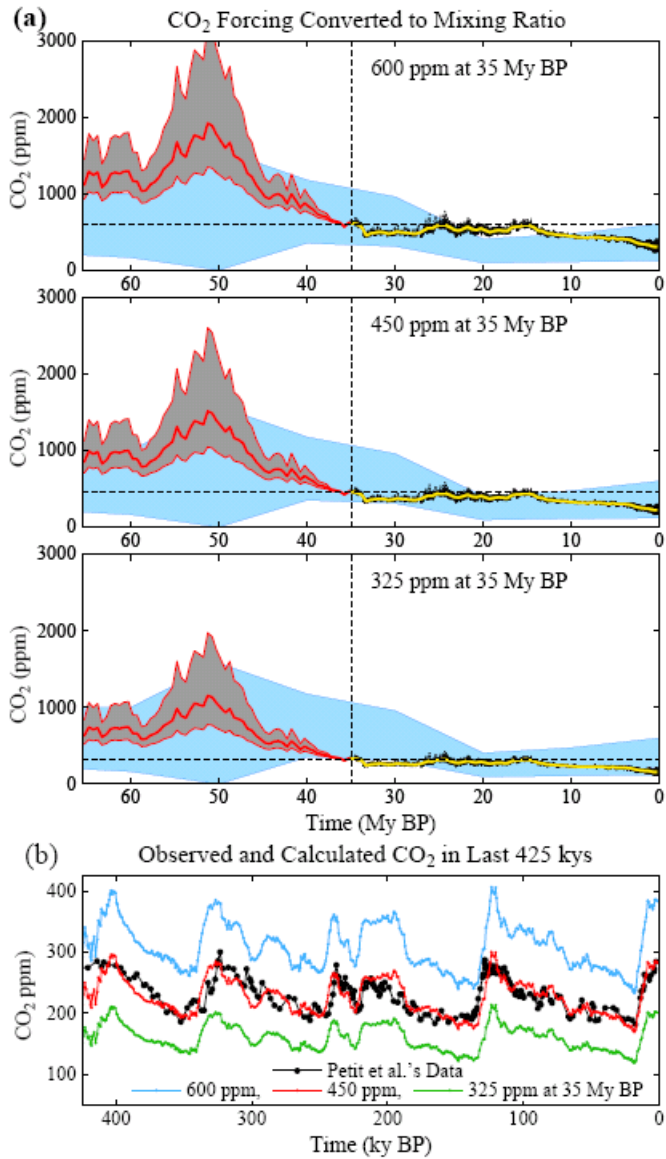

**Fig. (5). (a)** Simulated $CO_2$ amounts in the Cenozoic for three choices of $CO_2$ amount at 35 My (temporal resolution of black and colored curves as in Fig. **(3)**; blue region: multiple $CO_2$ proxy data, discussed with Fig. **S10**; gray region allows 50 percent uncertainty in ratio of global surface and deep ocean temperatures). **(b)** Expanded view of late Pleistocene, including precise ice core $CO_2$ measurements (black curve).

If we specify $CO_2$ at 35 My, the GHG forcing defines $CO_2$ at other times, assuming $CO_2$ provides 75% of the GHG forcing, as in the late Pleistocene. $CO_2$ ~450 ppm at 35 My keeps $CO_2$ in the range of early Cenozoic proxies (Fig. **5a**) and yields a good fit to the amplitude and mean $CO_2$ amount in the late Pleistocene (Fig. **5b**). A $CO_2$ threshold for Antarctic glaciation of ~500 ppm was previously inferred from proxy $CO_2$ data and a carbon cycle model [55].

Individual $CO_2$ proxies (Fig. **S10**) clarify limitations due to scatter among the measurements. Low $CO_2$ of some early Cenozoic proxies, if valid, would suggest higher climate sensitivity. However, in general the sensitivities inferred from the Cenozoic and Phanerozoic [56, 57, 58] agree well with our analysis, if we account for the ways in which sensitivity is defined and the periods emphasized in each empirical derivation (Table **S1**).

Our $CO_2$ estimate of ~450 ppm at 35 My (Fig. **5**) serves as a prediction to compare with new data on $CO_2$ amount. Model uncertainties (Fig. **S10**) include possible changes of non-$CO_2$ GHGs and the relation of $\Delta T_s$ to $\Delta T_{do}$. The model fails to account for cooling in the past 15 My if $CO_2$ increased, as several proxies suggest (Fig. **S10**). Changing ocean currents, such as the closing of the Isthmus of Panama, may have contributed to climate evolution, but models find little effect on temperature [59]. Non-$CO_2$ GHGs also could have played a role, because little forcing would have been needed to cause cooling due to the magnitude of late Cenozoic albedo feedback.

### 3.3. Implication

We infer from Cenozoic data that $CO_2$ was the dominant Cenozoic forcing, that $CO_2$ was ~450 ± 100 ppm when Antarctica glaciated, and that glaciation is reversible. Together these inferences have profound implications.

Consider three points marked in Fig. (**4**): point A at 35 My, just before Antarctica glaciated; point B at recent interglacial periods; point C at the depth of recent ice ages. Point B is about half way between A and C in global temperature (Fig. **3b**) and climate forcings (Fig. **4**). The GHG forcing from the deepest recent ice age to current interglacial warmth is ~3.5 W/m$^2$. Additional 4 W/m$^2$ forcing carries the planet, at equilibrium, to the ice-free state. Thus equilibrium climate sensitivity to GHG change, including the surface albedo change as a slow feedback, is almost as large between today and an ice-free world as between today and the ice ages.

The implication is that global climate sensitivity of 3°C for doubled $CO_2$, although valid for the idealized Charney definition of climate sensitivity, is a considerable understatement of expected equilibrium global warming in response to imposed doubled $CO_2$. Additional warming, due to slow climate feedbacks including loss of ice and spread of flora over the vast high-latitude land area in the Northern Hemisphere, approximately doubles equilibrium climate sensitivity.

Equilibrium sensitivity 6°C for doubled $CO_2$ is relevant to the case in which GHG changes are specified. That is appropriate to the anthropogenic case, provided the GHG amounts are estimated from carbon cycle models including climate feedbacks such as methane release from tundra and ocean sediments. The equilibrium sensitivity is even higher if the GHG feedback is included as part of the climate response, as is appropriate for analysis of the climate response to Earth orbital perturbations. The very high sensitivity with both albedo and GHG slow feedbacks included accounts for the huge magnitude of glacial-interglacial fluctuations in the Pleistocene (Fig. **3**) in response to small forcings (section 3 of Appendix).

Equilibrium climate response would not be reached in decades or even in a century, because surface warming is slowed by the inertia of the ocean (Fig. **S7**) and ice sheets. However, Earth's history suggests that positive feedbacks, especially surface albedo changes, can spur rapid global warmings, including sea level rise as fast as several meters per century [7]. Thus if humans push the climate system sufficiently far into disequilibrium, positive climate feedbacks may set in motion dramatic climate change and climate impacts that cannot be controlled.

### 4. ANTHROPOCENE ERA

Human-made global climate forcings now prevail over natural forcings (Fig. **2**). Earth may have entered the Anthropocene era [60, 61] 6-8 ky ago [62], but the net human-made forcing was small, perhaps slightly negative [7], prior to the industrial era. GHG forcing overwhelmed natural and negative human-made forcings only in the past quarter century (Fig. **2**).

Human-made climate change is delayed by ocean (Fig. **S7**) and ice sheet response times. Warming 'in the pipeline', mostly attributable to slow feedbacks, is now about 2°C (Fig. **2**). No additional forcing is required to raise global temperature to at least the level of the Pliocene, 2-3 million years ago, a degree of warming that would surely yield 'dangerous' climate impacts [5].

### 4.1. Tipping points

Realization that today's climate is far out of equilibrium with current climate forcings raises the specter of 'tipping points', the concept that climate can reach a point where, without additional forcing, rapid changes proceed practically out of our control [2, 7, 63, 64]. Arctic sea ice and the West Antarctic Ice Sheet are examples of potential tipping points. Arctic sea ice loss is magnified by the positive feedback of increased absorption of sunlight as global warming initiates sea ice retreat [65]. West Antarctic ice loss can be accelerated by several feedbacks, once ice loss is substantial [39].

We define: (1) the *tipping level*, the global climate forcing that, if long maintained, gives rise to a specific consequence, and (2) the *point of no return*, a climate state beyond which the consequence is inevitable, even if climate forcings are reduced. A point of no return can be avoided, even if the tipping level is temporarily exceeded. Ocean and ice sheet inertia permit overshoot, provided the climate forcing is returned below the tipping level before initiating irreversible dynamic change.

Points of no return are inherently difficult to define, because the dynamical problems are nonlinear. Existing models are more lethargic than the real world for phenomena now unfolding, including changes of sea ice [65], ice streams [66], ice shelves [36], and expansion of the subtropics [67, 68].

The tipping level is easier to assess, because the paleoclimate quasi-equilibrium response to known climate forcing is relevant. The tipping level is a measure of the long-term climate forcing that humanity must aim to stay beneath to avoid large climate impacts. The tipping level does not define the magnitude or period of tolerable overshoot. However, if overshoot is in place for centuries, the thermal perturbation will so penetrate the ocean [10] that recovery without dramatic effects, such as ice sheet disintegration, becomes unlikely.

**4.2. Target $CO_2$**

Combined, GHGs other than $CO_2$ cause climate forcing comparable to that of $CO_2$ [2, 6], but growth of non-$CO_2$ GHGs is falling below IPCC [2] scenarios. Thus total GHG climate forcing change is now determined mainly by $CO_2$ [69]. Coincidentally, $CO_2$ forcing is similar to the net human-made forcing, because non-$CO_2$ GHGs tend to offset negative aerosol forcing [2, 5].

Thus we take future $CO_2$ change as approximating the net human-made forcing change, with two caveats. First, special effort to reduce non-$CO_2$ GHGs could alleviate the $CO_2$ requirement, allowing up to about +25 ppm $CO_2$ for the same climate effect, while resurgent growth of non-$CO_2$ GHGs could reduce allowed $CO_2$ a similar amount [6]. Second, reduction of human-made aerosols, which have a net cooling effect, could force stricter GHG requirements. However, an emphasis on reducing black soot could largely off-set reductions of high albedo aerosols [20].

Our estimated history of $CO_2$ through the Cenozoic Era provides a sobering perspective for assessing an appropriate target for future $CO_2$ levels. A $CO_2$ amount of order 450 ppm or larger, if long maintained, would push Earth toward the ice-free state. Although ocean and ice sheet inertia limit the rate of climate change, such a $CO_2$ level likely would cause the passing of climate tipping points and initiate dynamic responses that could be out of humanity's control.

The climate system, because of its inertia, has not yet fully responded to the recent increase of human-made climate forcings [5]. Yet climate impacts are already occurring that allow us to make an initial estimate for a target atmospheric $CO_2$ level. No doubt the target will need to be adjusted as climate data and knowledge improve, but the urgency and difficulty of reducing the human-made forcing will be less, and more likely manageable, if excess forcing is limited soon.

Civilization is adapted to climate zones of the Holocene. Theory and models indicate that subtropical regions expand poleward with global warming [2, 67]. Data reveal a 4-degree latitudinal shift already [68], larger than model predictions, yielding increased aridity in southern United States [70, 71], the Mediterranean region, Australia and parts of Africa. Impacts of this climate shift [72] support the conclusion that 385 ppm $CO_2$ is already deleterious.

Alpine glaciers are in near-global retreat [72, 73]. After a one-time added flush of fresh water, glacier demise will yield summers and autumns of frequently dry rivers, including rivers originating in the Himalayas, Andes and Rocky Mountains that now supply water to hundreds of millions of people. Present glacier retreat, and warming in the pipeline, indicate that 385 ppm $CO_2$ is already a threat.

Equilibrium sea level rise for today's 385 ppm $CO_2$ is at least several meters, judging from paleoclimate history [19, 32-34]. Accelerating mass losses from Greenland [74] and West Antarctica [75] heighten concerns about ice sheet stability. An initial $CO_2$ target of 350 ppm, to be reassessed as effects on ice sheet mass balance are observed, is suggested.

Stabilization of Arctic sea ice cover requires, to first approximation, restoration of planetary energy balance. Climate models driven by known forcings yield a present planetary energy imbalance of +0.5-1 $W/m^2$ [5]. Observed heat increase in the upper 700 m of the ocean [76] confirms the planetary energy imbalance, but observations of the entire ocean are needed for quantification. $CO_2$ amount must be reduced to 325-355 ppm to increase outgoing flux 0.5-1 $W/m^2$, if other forcings are unchanged. A further imbalance reduction, and thus $CO_2$ ~300-325 ppm, may be needed to restore sea ice to its area of 25 years ago.

Coral reefs are suffering from multiple stresses, with ocean acidification and ocean warming principal among them [77]. Given additional warming 'in-the-pipeline', 385 ppm $CO_2$ is already deleterious. A 300-350 ppm $CO_2$ target would significantly relieve both of these stresses.

### 4.3. $CO_2$ scenarios

A large fraction of fossil fuel $CO_2$ emissions stays in the air a long time, one-quarter remaining airborne for several centuries [11, 78, 79]. Thus moderate delay of fossil fuel use will not appreciably reduce long-term human-made climate change. Preservation of a climate resembling that to which humanity is accustomed, the climate of the Holocene, requires that most remaining fossil fuel carbon is never emitted to the atmosphere.

Coal is the largest reservoir of conventional fossil fuels (Fig. **S12**), exceeding combined reserves of oil and gas [2, 79]. The only realistic way to sharply curtail $CO_2$ emissions is to phase out coal use except where $CO_2$ is captured and sequestered.

Phase-out of coal emissions by 2030 (Fig. **6**) keeps maximum $CO_2$ close to 400 ppm, depending on oil and gas reserves and reserve growth. IPCC reserves assume that half of readily extractable oil has already been used (Figs. **6, S12**). EIA [80] estimates (Fig. **S12**) have larger reserves and reserve growth. Even if EIA estimates are accurate, the IPCC case remains valid if the most difficult to extract oil and gas is left in the ground, via a rising price on carbon emissions that discourages remote exploration and environmental regulations that place some areas off-limit. If IPCC gas reserves (Fig. **S12**) are underestimated, the IPCC case in Fig. (**6**) remains valid if the additional gas reserves are used at facilities where $CO_2$ is captured.

However, even with phase-out of coal emissions and assuming IPCC oil and gas reserves, $CO_2$ would remain above 350 ppm for more than two centuries. Ongoing Arctic and ice sheet changes, examples of rapid paleoclimate change, and other criteria cited above all drive us to consider scenarios that bring $CO_2$ more rapidly back to 350 ppm or less.

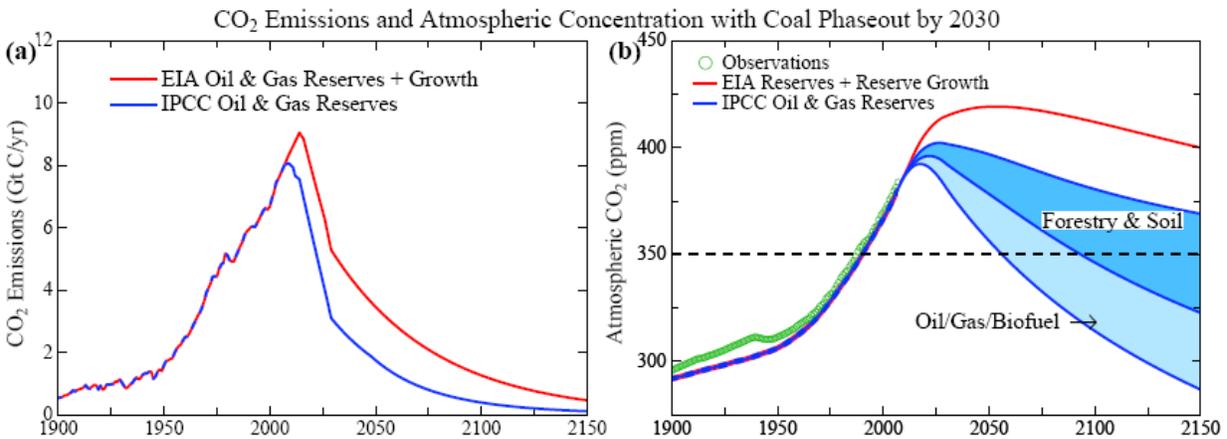

**Fig. (6). (a)** Fossil fuel $CO_2$ emissions with coal phase-out by 2030 based on IPCC [2] and EIA [80] estimated fossil fuel reserves. **(b)** Resulting atmospheric $CO_2$ based on use of a dynamic-sink pulse response function representation of the Bern carbon cycle model [78, 79].

### 4.4. Policy relevance

Desire to reduce airborne $CO_2$ raises the question of whether $CO_2$ could be drawn from the air artificially. There are no large-scale technologies for $CO_2$ air capture now, but with strong research and development support and industrial-scale pilot projects sustained over decades it may be possible to achieve costs ~$200/tC [81] or perhaps less [82]. At $200/tC, the cost of removing 50 ppm of $CO_2$ is ~$20 trillion.

Improved agricultural and forestry practices offer a more natural way to draw down $CO_2$. Deforestation contributed a net emission of 60±30 ppm over the past few hundred years, of which ~20 ppm $CO_2$ remains in the air today [2, 83, Figs. **(S12), (S14)**]. Reforestation could absorb a substantial fraction of the 60±30 ppm net deforestation emission.

Carbon sequestration in soil also has significant potential. Biochar, produced in pyrolysis of residues from crops, forestry, and animal wastes, can be used to restore soil fertility while storing carbon for centuries to millennia [84]. Biochar helps soil retain nutrients and fertilizers, reducing emissions of GHGs such as $N_2O$ [85]. Replacing slash-and-burn agriculture with slash-and-char and use of agricultural and forestry wastes for biochar production could provide a $CO_2$ drawdown of ~8 ppm or more in half a century [85].

In the Appendix we define a forest/soil drawdown scenario that reaches 50 ppm by 2150 (Fig. **6b**). This scenario returns $CO_2$ below 350 ppm late this century, after about 100 years above that level.

More rapid drawdown could be provided by $CO_2$ capture at power plants fueled by gas and biofuels [86]. Low-input high-diversity biofuels grown on degraded or marginal lands, with associated biochar production, could accelerate $CO_2$ drawdown, but the nature of a biofuel approach must be carefully designed [85, 87-89].

A rising price on carbon emissions and payment for carbon sequestration is surely needed to make drawdown of airborne $CO_2$ a reality. A 50 ppm drawdown via agricultural and forestry practices seems plausible. But if most of the $CO_2$ in coal is put into the air, no such "natural" drawdown of $CO_2$ to 350 ppm is feasible. Indeed, if the world continues on a business-as-usual path for even another decade without initiating phase-out of unconstrained coal use, prospects for avoiding a dangerously large, extended overshoot of the 350 ppm level will be dim.

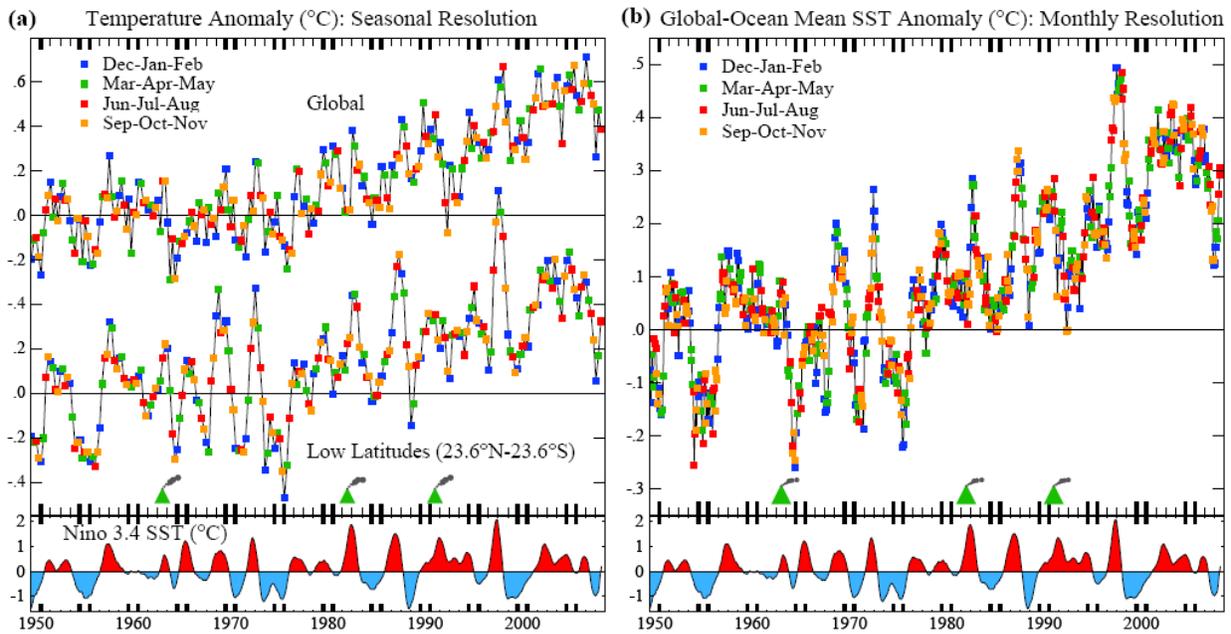

**Fig. (7). (a)** Seasonal-mean global and low-latitude surface temperature anomalies relative to 1951-1980, an update of [92], **(b)** global-ocean-mean sea surface temperature anomaly at monthly resolution. The Nino 3.4 Index, the temperature anomaly (12-month running mean) in a small part of the tropical Pacific Ocean [93], is a measure of ENSO, a basin-wide sloshing of the tropical Pacific Ocean [94]. Green triangles show major volcanic eruptions.

### 4.5. Caveats: climate variability, climate models, and uncertainties

Climate has great variability, much of which is unforced and unpredictable [2, 90]. This fact raises a practical issue: what is the chance that climate variations, e.g., a temporary cooling trend, will affect public recognition of climate change, making it difficult to implement mitigation policies? Also what are the greatest uncertainties in the expectation of a continued global warming trend? And what are the impacts of climate model limitations, given the inability of models to realistically simulate many aspects of climate change and climate processes?

The atmosphere and ocean exhibit coupled nonlinear chaotic variability that cascades to all time scales [91]. Variability is so large that the significance of recent decadal global temperature change (Fig. 7a) would be very limited, if the data were considered simply as a time series, without further information. However, other knowledge includes information on the causes of some of the temperature variability, the planet's energy imbalance, and global climate forcings.

The El Nino Southern Oscillation (ENSO) [94] accounts for most low latitude temperature variability and much of the global variability. The global impact of ENSO is coherent from month to month, as shown by the global-ocean-mean SST (Fig. 7b), for which the ocean's thermal inertia minimizes the effect of weather noise. The cool anomaly of 2008 coincides with an ENSO minimum and does not imply a change of decadal temperature trend.

Decadal time scale variability, such as predicted weakening of the Atlantic overturning circulation [95], could interrupt global warming, as discussed in section 18 of the Appendix. But the impact of regional dynamical effects on global temperature is opposed by the planet's energy imbalance [96], a product of the climate system's thermal inertia, which is confirmed by increasing ocean heat storage [97]. This energy imbalance makes decadal interruption of global warming, in the absence of a negative climate forcing, improbable [96].

Volcanoes and the sun can cause significant negative forcings. However, even if the solar irradiance remained at its value in the current solar minimum, this reduced forcing would be offset by increasing $CO_2$ within seven years (Appendix section 18). Human-made aerosols cause a greater negative forcing, both directly and through their effects on clouds. The first satellite observations of aerosols and clouds with accuracy sufficient to quantify this forcing are planned to begin in 2009 [98], but most analysts anticipate that human-made aerosols will decrease in the future, rather than increase further.

Climate models have many deficiencies in their abilities to simulate climate change [2]. However, model uncertainties cut both ways: it is at least as likely that models underestimate effects of human-made GHGs as overestimate them (Appendix section 18). Model deficiencies in evaluating tipping points, the possibility that rapid changes can occur without additional climate forcing [63, 64], are of special concern. Loss of Arctic sea ice, for example, has proceeded more rapidly than predicted by climate models [99]. There are reasons to expect that other nonlinear problems, such as ice sheet disintegration and extinction of interdependent species and ecosystems, also have the potential for rapid change [39, 63, 64].

## 5. SUMMARY

Humanity today, collectively, must face the uncomfortable fact that industrial civilization itself has become the principal driver of global climate. If we stay our present course, using fossil fuels to feed a growing appetite for energy-intensive life styles, we will soon leave the climate of the Holocene, the world of prior human history. The eventual response to doubling pre-industrial atmospheric $CO_2$ likely would be a nearly ice-free planet, preceded by a period of chaotic change with continually changing shorelines.

Humanity's task of moderating human-caused global climate change is urgent. Ocean and ice sheet inertias provide a buffer delaying full response by centuries, but there is a danger that human-made forcings could drive the climate system beyond tipping points such that change proceeds out of our control. The time available to reduce the human-made forcing is uncertain, because models of the global system and critical components such as ice sheets are inadequate. However, climate response time is surely less than the atmospheric lifetime of the human-caused perturbation of $CO_2$. Thus remaining fossil fuel reserves should not be exploited without a plan for retrieval and disposal of resulting atmospheric $CO_2$.

Paleoclimate evidence and ongoing global changes imply that today's $CO_2$, about 385 ppm, is already too high to maintain the climate to which humanity, wildlife, and the rest of the biosphere are adapted. Realization that we must reduce the current $CO_2$ amount has a bright side: effects that had begun to seem inevitable, including impacts of ocean acidification, loss of fresh water supplies, and shifting of climatic zones, may be averted by the necessity of finding an energy course beyond fossil fuels sooner than would otherwise have occurred.

We suggest an initial objective of reducing atmospheric $CO_2$ to 350 ppm, with the target to be adjusted as scientific understanding and empirical evidence of climate effects accumulate. Although a case already could be made that the eventual target probably needs to be lower, the 350 ppm target is sufficient to qualitatively change the discussion and drive fundamental changes in energy policy. Limited opportunities for reduction of non-$CO_2$ human-caused forcings are important to pursue but do not alter the initial 350 ppm $CO_2$ target. This target must be pursued on a timescale of decades, as paleoclimate and ongoing changes, and the ocean response time, suggest that it would be foolhardy to allow $CO_2$ to stay in the dangerous zone for centuries.

A practical global strategy almost surely requires a rising global price on $CO_2$ emissions and phase-out of coal use except for cases where the $CO_2$ is captured and sequestered. The carbon price should eliminate use of unconventional fossil fuels, unless, as is unlikely, the $CO_2$ can be captured. A reward system for improved agricultural and forestry practices that sequester carbon could remove the current $CO_2$ overshoot. With simultaneous policies to reduce non-$CO_2$ greenhouse gases, it appears still feasible to avert catastrophic climate change.

Present policies, with continued construction of coal-fired power plants without $CO_2$ capture, suggest that decision-makers do not appreciate the gravity of the situation. We must begin to move now toward the era beyond fossil fuels. Continued growth of greenhouse gas emissions, for just another decade, practically eliminates the possibility of near-term return of atmospheric composition beneath the tipping level for catastrophic effects.

The most difficult task, phase-out over the next 20-25 years of coal use that does not capture $CO_2$, is Herculean, yet feasible when compared with the efforts that went into World War II. The stakes, for all life on the planet, surpass those of any previous crisis. The greatest danger is continued ignorance and denial, which could make tragic consequences unavoidable.

## ACKNOWLEDGMENTS

We thank H. Harvey and Hewlett Foundation, G. Lenfest, the Rockefeller Family Foundation, and NASA program managers D. Anderson and J. Kaye for research support, an anonymous reviewer, S. Baum, B. Brook, P. Essunger, K. Farnish, Q. Fu, L.D. Harvey, I. Horovitz, R. Keeling, C. Kutscher, J. Leventhal, C. McGrath, T. Noerpel, P. Read, J. Romm, D. Sanborn, S. Schwartz, J. Severinghaus, K. Ward and S. Weart for comments on a draft manuscript, G. Russell for the computations in Fig. S3, and T. Conway and P. Tans of NOAA Earth System Research Laboratory and R. Andres, T. Boden and G. Marland of DOE CDIAC for data.

## REFERENCES

[1] Framework Convention on Climate Change, United Nations, 1992; http://www.unfccc.int/.
[2] Intergovernmental Panel on Climate Change (IPCC), Climate Change 2007, Solomon S, Dahe Q, Manning M, *et al.* (eds), Cambridge Univ Press: New York, 2007; pp. 996.
[3] Mastrandrea MD, Schneider SH. Probabilistic integrated assessment of "dangerous" climate change. Science 2004; 304: 571-575.
[4] European Council, Climate change strategies. 2005; http://register.consilium.europa.eu/pdf/en/05/st07/st07242.en05.pdf
[5] Hansen J, Sato M, Ruedy, *et al.* Dangerous human-made interference with climate: a GISS modelE study. Atmos Chem Phys 2007; 7: 2287-2312.
[6] Hansen J, Sato M. Greenhouse gas growth rates. Proc Natl Acad Sci 2004; 101: 16109-16114.
[7] Hansen J, Sato M, Kharecha P, Russell G, Lea D W and Siddall M. Climate change and trace gases. Phil Trans R Soc A 2007; 365: 1925-1954.
[8] Hansen J, Nazarenko L, Ruedy R, *et al.* Earth's energy imbalance: Confirmation and implications. Science 2005; 308: 1431-1435.
[9] Harvey LDD. Dangerous anthropogenic interference, dangerous climatic change, and harmful climatic change: non-trivial distinctions with significant policy implications. Clim Change 2007; 82: 1-25.
[10] Matthews HD, Caldeira K. Stabilizing climate requires near-zero emissions. Geophys Res Lett 2008; 35: L04705.
[11] Archer D. Fate of fossil fuel $CO_2$ in geologic time. J Geophys Res 2005; 110:C09S05.
[12] Hansen J, Sato M, Ruedy R, *et al.* Efficacy of climate forcings. J Geophys Res 2005; 110: D18104.
[13] Charney J. Carbon Dioxide and Climate: A Scientific Assessment. National Academy of Sciences Press: Washington DC, 1979; pp. 33.
[14] Hansen J, Lacis A, Rind D, Russell G, Stone P, Fung I, Ruedy R, Lerner J. Climate sensitivity: Analysis of feedback mechanisms. In Climate Processes and Climate Sensitivity, Geophys. Monogr. Ser. 29 (eds. Hansen JE, Takahashi T), American Geophysical Union: Washington, D.C., 1984; pp. 130-163.
[15] Braconnot P, Otto-Bliesner BL, Harrison S, *et al.*, Results of PMIP2 coupled simulations of the Mid-Holocene and Last Glacial Maximum – Part 1: experiments and large-scale features. Clim. Past 2007; 3: 261-277.
[16] Farrera I, Harrison SP, Prentice IC, *et al.* Tropical climates at the last glacial maximum: a new synthesis of terrestrial paeleoclimate data. I. Vegetation, lake-levels and geochemistry. Clim Dyn 1999; 15: 823-856
[17] Petit JR, Jouzel J, Raynaud D, Barkov NI, Barnola JM, Basile I, Bender M, Chappellaz J, Davis J, Delaygue G, *et al.* 420,000 years of climate and atmospheric history revealed by the Vostok deep Antarctic ice core. Nature 1999; 399: 429-436.
[18] Vimeux F, Cuffey KM, Jouzel J. New insights into Southern Hemisphere temperature changes from Vostok ice cores using deuterium excess correction. Earth Planet Sci Lett 2002; 203: 829-843.

[19] Siddall M, Rohling EJ, Almogi-Labin A, Hemleben Ch, Meischner D, Schmelzer I, Smeed DA. Sea-level fluctuations during the last glacial cycle. Nature 2003; 423: 853-858.
[20] Hansen J, Sato M, Ruedy R, Lacis A, Oinas V. Global warming in the twenty-first century: An alternative scenario. Proc Natl Acad Sci. 2000; 97: 9875-9880.
[21] Masson-Delmotte V, Kageyama M, Braconnot P. Past and future polar amplification of climate change: climate model intercomparisons and ice-core constraints. Clim Dyn 2006; 26: 513-529.
[22] EPICA community members. One-to-one coupling of glacial climate variability in Greenland and Antarctica. Nature 2006; 444: 195-198.
[23] Caillon N, Severinghaus JP, Jouzel J, Barnola JM, Kang J, Lipenkov VY. Timing of atmospheric $CO_2$ and Antarctic temperature changes across Termination III. Science 2003; 299: 1728-1731.
[24] Mudelsee M. The phase relations among atmospheric CO2 content, temperature and global ice volume over the past 420 ka. Quat Sci Rev 2001; 20: 583-589.
[25] Hays JD, Imbrie J, Shackleton NJ. Variations in the Earth's orbit: pacemaker of the ice ages. Science 1976; 194: 1121-1132.
[26] Zachos J, Pagani M, Sloan L, Thomas E, Billups K. Trends, rhythms, and aberrations in global climate 65 Ma to present. Science 2001; 292: 686-693.
[27] Kohler P, Fischer H. Simulating low frequency changes in atmospheric $CO_2$ during the last 740 000 years. Clim Past 2006; 2: 57-78.
[28] Siegenthaler U, Stocker TF, Monnin E, Luthi D, Schwander J, Stauffer B, Raynaud D, Barnola JM, Fischer H, Masson-Delmotte V, Jouzel J. Stable carbon cycle – climate relationship during the late Pleistocene. Science 2005; 310: 1313-1317.
[29] Archer D. Methane hydrate stability and anthropogenic climate change. Biogeosci 2007; 4: 521-544.
[30] Berner RA. The Phanerozoic Carbon Cycle: $CO_2$ and $O_2$; Oxford Univ Press: New York, 2004; pp. 150.
[31] Hansen J, Russell G, Lacis A, *et al.*, Climate response times: Dependence on climate sensitivity and ocean mixing. Science 1985; 229: 857-859.
[32] Thompson WG, Goldstein SL. Open-system coral ages reveal persistent suborbital sea-level cycles. Science 2005; 308: 401-404.
[33] Hearty PJ, Hollin JT, Neumann AC, O'Leary MJ, McCulloch M. Global sea-level fluctuations during the last interglaciation (MIS 5e). Quat Sci Rev 2007; 26: 2090-2112.
[34] Rohling EJ, Grant K, Hemleben Ch, Siddall M, Hoogakker BAA, Bolshaw M, Kucera M. High rates of sea-level rise during the last interglacial period. Nature Geosci 2008; 1: 38-42.
[35] Tedesco M. Snowmelt detection over the Greenland ice sheet from SSM/I brightness temperature daily variations. Geophys Res Lett 2007; 34: L02504, 1-6.
[36] Rignot E, Jacobs SS. Rapid bottom melting widespread near Antarctic ice sheet grounding lines. Science 2002; 296: 2020-2023.
[37] Zwally HJ, Abdalati W, Herring T, Larson K, Saba J, Steffen K. Surface melt-induced acceleration of Greenland ice-sheet flow. Science 2002; 297: 218-222.
[38] Chen JL, Wilson CR, Tapley BD. Satellite gravity measurements confirm accelerated melting of Greenland Ice Sheet. Science 2006; 313: 1958-1960.
[39] Hansen J. A slippery slope: how much global warming constitutes "dangerous anthropogenic interference"? Climatic Change 2005; 68: 269-279.
[40] DeConto RM, Pollard D. Rapid Cenozoic glaciation of Antarctica induced by declining atmospheric $CO_2$. Nature 2003; 421: 245-249.
[41] Zanazzi A, Kohn MJ, MacFadden BJ, Terry DO. Large temperature drop across the Eocene-Oligocene transition in central North America. Nature 2007; 445: 639-642.
[42] Dupont-Nivet G, Krijgsman W, Langereis CG, Abeld HA, Dai S, Fang X. Tibetan plateau aridification linked to global cooling at the Eocene–Oligocene transition. Nature 2007; 445: 635-638.
[43] Sackmann IJ, Boothroyd AI, Kraemer KE. Our sun III Present and future. Astrophys J 1993; 418: 457-468.
[44] Pagani M, Zachos J, Freeman KH, Bohaty S, Tipple B. Marked change in atmospheric carbon dioxide concentrations during the Oligocene. Science 2005; 309: 600-603.
[45] Bartdorff O, Wallmann K, Latif M, Semenov V. Phanerozoic evolution of atmospheric methane. Global Biogeochem Cycles 2008; 22: GB1008.
[46] Beerling D, Berner RA, Mackenzie FT, Harfoot MB, Pyle JA. Methane and the $CH_4$ greenhouse during the past 400 million years. Am J Sci 2008; in press.
[47] Edmond JM, Huh Y. Non-steady state carbonate recycling and implications for the evolution of atmospheric $P_{CO_2}$. Earth Planet. Sci Lett 2003; 216: 125-139.
[48] Staudigel H, Hart SR, Schmincke H-U, Smith BM. Cretaceous ocean crust at DSDP Sites 417 and 418: Carbon uptake from weathering versus loss by magmatic outgassing. Geochim Cosmochim Acta 1989; 53: 3091-3094.
[49] Berner R, Caldeira K. The need for mass balance and feedback in the geochemical carbon cycle. Geology 1997; 25: 955-956.

[50] Kumar P, Yuan X, Kumar MR, Kind R, Li X, Chadha RK. The rapid drift of the Indian tectonic plate. Nature 2007; 449: 894-897.
[51] Raymo ME, Ruddiman WF. Tectonic forcing of late Cenozoic climate. Nature 1992; 359, 117-122.
[52] Zeebe RE, Caldeira K. Close mass balance of long-term carbon fluxes from ice-core $CO_2$ and ocean chemistry records. Nature Geosci 2008; 1: 312-315.
[53] Patriat P, Sloan H, Sauter D. From slow to ultraslow: a previously undetected event at the Southwest Indian Ridge at ca. 24 Ma. Geology 2008; 36: 207-210.
[54] Joshi MM, Gregory JM, Webb MJ, Sexton DMH, Johns TC. Mechanisms for the land/sea warming contrast exhibited by simulations of climate change. Clim Dyn 2008; 30: 455-465.
[55] Royer DL. $CO_2$-forced climate thresholds during the Phanerozoic. Geochim Cosmochim Acta 2006; 70: 5665-5675.
[56] Royer DL, Berner RA, Park J. Climate sensitivity constrained by $CO_2$ concentrations over the past 420 million years. Nature 2007; 446: 530-532.
[57] Higgins JA, Schrag DP. Beyond methane: Towards a theory for Paleocene-Eocene Thermal Maximum. Earth Planet Sci Lett 2006; 245: 523-537.
[58] Pagani M, Caldeira K, Archer D, Zachos JC. An ancient carbon mystery. Science 2006; 314: 1556-11557.
[59] Lunt DJ, Valdes PJ, Haywood A, Rutt IC. Closure of the Panama Seaway during the Pliocene: implications for climate and Northern Hemisphere glaciation. Clim. Dyn 2008; 30: 1-18.
[60] Crutzen PJ, Stoermer EF. The "Anthropocene". Global Change Newsletter 2000; 41: 12-13.
[61] Zalasiewicz J, Williams M, Smith A, *et al.* Are we now living in the Anthropocene? GSA Today 2008; 18: 4-8.
[62] Ruddiman WF. The anthropogenic greenhouse era began thousands of years ago. Clim. Change 2003; 61: 261-293.
[63] Hansen J. Tipping point: perspective of a climatologist. In *State of the Wild: A Global Portrait of Wildlife, Wildlands, and Oceans.* 2008. W. Woods, Ed. Wildlife Conservation Society/Island Press, pp. 6-15.
[64] Lenton TM, Held H, Kriegler E, Hall JW, Lucht W, Rahmstorf S, Schellnhuber HJ. Tipping elements in the Earth's climate system. Proc Natl Acad Sci 2008; 105: 1786-1793.
[65] Stroeve J, Serreze M, Drobot S, Gearheard S, Holland M, Maslanik J, Meier W, Scambos T. Arctic sea ice extent plummets in 2007. Eos Trans, AGU 2008; 89(2): 13.
[66] Howat IM, Joughin I, Scambos TA. Rapid changes in ice discharge from Greenland outlet glaciers. Science 2007; 315: 1559-1561.
[67] Held IM, Soden BJ. Robust responses of the hydrological cycle to global warming. J Climate 2006; 19: 5686-5699.
[68] Seidel DJ, Randel WJ. Variability and trends in the global tropopause estimated from radiosonde data. J Geophys Res 2006; 111: D21101.
[69] Hansen J, Sato M. Global warming: East-West connections. Open Environ J 2008 (to be submitted).
[70] Barnett TP, Pierce DW, Hidalgo HG, *et al*. Human-induced changes in the hydrology of the Western United States. Science 2008; 319: 1080-1083.
[71] Levi BG. Trends in the hydrology of the western US bear the imprint of manmade climate change. Phys Today 2008; April: 16-18.
[72] Intergovernmental Panel on Climate Change (IPCC), Impacts, Adaptation and Vulnerability, M. Parry *et al.*, (Eds.) Cambridge Univ. Press: New York, 2007; pp. 978.
[73] Barnett TP, Adam JC, Lettenmaler DP. Potential impacts of a warming climate on water availability in snow-dominated regions. Nature 2005; 438: 303-309.
[74] Steffen K *et al.*, Chap. 2 in Abrupt Climate Change, U.S. Climate Change Science Program, SAP-3.4 2008; (in press).
[75] Rignot E, Bamber JL, van den Broeke MR, Davis C, Li Y, van de Berg WJ, van Meijgaard E. Recent Antarctic ice mass loss from radar interferometry and regional climate modeling. Nature Geoscience 2008; 1, 106-110.
[76] Domingues CM, Church JA, White NJ, Gleckler PJ, Wijffels, Barker PM, Dunn JR. Rapid upper-ocean warming helps explain multi-decadal sea-level rise. Nautre 2008 (in press).
[77] Stone R. A world without corals? Science 2007; 316: 678-681.
[78] Joos F, Bruno M, Fink R, Stocker TF, Siegenthaler U, Le Quere C, Sarmiento JL. An efficient and accurate representation of complex oceanic and biospheric models of anthropogenic carbon uptake. Tellus B 1996; 48: 397-417.
[79] Kharecha P, Hansen J. Implications of "peak oil" for atmospheric CO2 and climate. Global Biogeochem Cycles 2008; 22: GB3012.
[80] Energy Information Administration (EIA), U.S. DOE, International Energy Outlook 2006, http://www.eia.doe.gov/oiaf/archive/ieo06/index.html.
[81] Keith DW, Ha-Duong M, Stolaroff JK. Climate strategy with $CO_2$ capture from the air. Clim Change 2006; 74: 17-45.
[82] Lackner KS. A Guide to $CO_2$ Sequestration. Science 2003; 300: 1677-1678.

| | |
|---|---|
| [83] | Houghton RA. Revised estimates of the annual net flux of carbon to the atmosphere from changes in land use and land management 1850-2000. Tellus B 2003; 55: 378-390. |
| [84] | Lehmann J. A handful of carbon. Nature 2007; 447: 143-144. |
| [85] | Lehmann J, Gaunt J, Rondon M. Bio-char sequestration in terrestrial ecosystems – a review. Mitigation and Adaptation Strategies for Global Change 2006; 11: 403-427. |
| [86] | Hansen J. Congressional Testimony, 2007; http://arxiv.org/abs/0706.3720v1. |
| [87] | Tilman D, Hill J, Lehman C. Carbon-negative biofuels from low-input high-diversity grassland biomass. Science 2006; 314: 1598-1600. |
| [88] | Fargione J, Hill J, Tilman D, Polasky S, Hawthorne P. Land clearing and the biofuel carbon debt. Science 2008; 319: 1235-1238. |
| [89] | Searchinger T, Heimlich R, Houghton RA, Dong F, Elobeid A, Fabiosa J, Tokgoz S, Hayes D, Yu T-H. Use of U.S. croplands for biofuels increases greenhouse gases through emissions from land-use change. Science 2008; 319: 1238-1240. |
| [90] | Palmer TN. Nonlinear dynamics and climate change: Rossby's legacy. Bull Amer Meteorol Soc 1998; 79: 1411-1423. |
| [91] | Hasselmann K. Ocean circulation and climate change. Tellus B 2002; 43: 82-103. |
| [92] | Hansen J, Ruedy R, Glascoe J, Sato M. GISS analysis of surface temperature change. J Geophys Res 1999; 104: 30997-31022. |
| [93] | NOAA National Weather Service, Climate Prediction Center. 2008; http://www.cpc.ncep.noaa.gov/data/indices/sstoi.indices |
| [94] | Cane, MA. El Nino. Ann Rev Earth Planet Sci 1986; 14: 43-70. |
| [95] | Keenlyside NS, Latif M, Jungclaus J, Kornblueh L, Roeckner E. Advancing decadal-scale climate prediction in the North Atlantic sector. Nature 2008; 453: 84-88. |
| [96] | Hansen J, Sato M, Ruedy R, *et al.* Forcings and chaos in interannual to decadal climate change. J Geophys Res 1997; 102: 25679-25720. |
| [97] | Domingues CM, Church JA, White NJ, Gleckler PJ, Wijffels SE, Barker PM, Dunn JR. Improved estimates of upper-ocean warming and multi-decadal sea-level rise. Nature 2008; 453: 1090-1093. |
| [98] | Mishchenko MI, Cairns B, Kopp G, Schueler BA, Fafaul BA, Hansen JE, Hooker RJ, Itchkawich T, Maring HB, Travis LD. Precise and accurate monitoring of terrestrial aerosols and total solar irradiance: introducing the Glory mission. Bull Amer Meteorol Soc 2007; 88: 677-691. |
| [99] | Lindsay RW, Zhang J. The Thinning of Arctic Sea Ice, 1988–2003: Have we passed a tipping point? J Climate 2005; 18: 4879-4894. |